# Development of FEB Configuration Test Board for ATLAS NSW Upgrade

Houbing Lu, Feng Li, Peng Miao, Kun Hu, Zhilei Zhang, Rongqi Sun, Qingli Ma, and Ge Jin

*Abstract*–The FEB(front end board) configuration test board is developed aiming at meeting the requirement of testing the new generation ASIC(application-specific integrated circuit) chips and its configuration system for ATLAS NSW(New Small Wheel) upgrade, In this paper, some functions are developed in terms of the configurations of the key chips on the FEB－VMM3 and TDS2 using GBT-SCA. Additionally, a flexible communication protocol is designed, verifying the whole data link. It provides technical reference for prototype FEB key chip configuration and data readout, as well as the final system configuration.

## I. INTRODUCTION

ATLAS[1] trigger system is responsible for recording the signal of interesting events, and it can eliminate most of the background events so as to greatly reduce the write rate, the ATLAS uses multi-layer trigger mode to build highly selective and efficient trigger system, so as to achieve efficient reception of collision events, and reduce overall data rate. The ATLAS hardware gets the trigger information from the CALO detectors and the Muon detectors, and reduces the input data rate from the 40MHz to about 75kHz. The NSW trigger is a complex and highly configurable system, each FEB has a GBT-SCA[2] chip on it specially designed for FEB mode configuration. This paper mainly discusses the development process of FEB configuration test board, which completes the verification of FEB system configuration function, masters the use of GBT-SCA, configures VMM3[3-6] and TDS2[7] chips on FEB, and studies the configuration process of NSW.

## II. ARCHITECTURE OF NSW CONFIGURATION SYSTEM

The devices that NSW need to be configured include VMM, ROC, GBTx, GBTLD and TDS. The configuration of the front-end hardware is completed through the GBT-SCA configuration chip, and the configuration data is generated by the back-end configuration system. The front-end configuration interface signal relationship is shown in Figure 1.

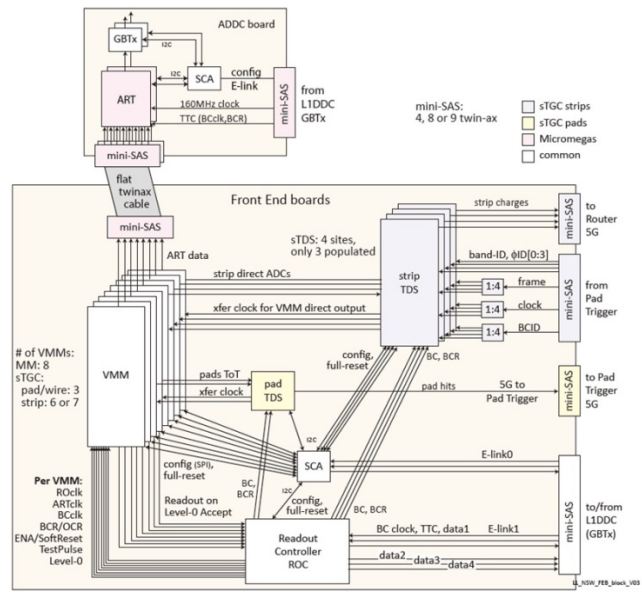

Fig. 1. Diagram of NSW Configuration.

The FEB configuration signals are transmitted to FEB by the control room through L1DDC (L1 Data Driver Card), the communication between L1DDC and FEB adopts E-link communication protocol, and the mini-SAS connector is used. The configuration functions that NSW needs to be completed are as follows:

(1) Each FEB for Micromegas and sTGC detector will be configured, and each board should be equipped with GBT-SCA ASIC, this means that for each FEB, a dedicated E-link configuration interface will be used, GBT-SCA has two E-link configuration interfaces, one for master and one for standby. The sTGC detector has a total of 1536 FEBs, and the Micromegas has a total of 4096 FEBs.

(2) We need to configure each VMM and TDS on the board. VMM configuration data is completed by the SPI bus, the GBT-SCA has 8 SPI interfaces, there are up to 7 VMM chips on one FEB. TDS configuration data is completed by the I2C bus, the GBT-SCA has 16 I2C interfaces, there are up to 4 TDS chips on one FEB, so placing a piece of GBT-SCA on one FEB can meet the configuration requirement of VMMs and TDSs.

(3) The GBT-SCA chip is integrated with an ADC, there are 32 external input interfaces. we can select the ADC channel sampling through an internal multiplexer, and the ADC input voltage ranges from 0 to 1V. The voltage sampling monitoring

Manuscript received May 30, 2018. This work is supported by the National Natural Science Foundation of China under grant number 11461141010, 11375179, and in part by "the Fundamental Research Funds for the Central Universities" under grant No. WK2360000005, and "Research Plan Project of the College" under grant No. ZK17-03-30.

Houbing Lu and Qingli Ma are with the College of Electronic Engineering, National University of Defense and Technology, Hefei, Anhui 230037, P.R. of China (e-mail: luhb@mail.ustc.edu.cn, maql@.ustc.edu.cn).

Kun. Hu was with State Key Laboratory of Particle Detection and Electronics, University of Science and Technology of China, Hefei, Anhui, 230026, China. He is now with Department of Radiation Oncology, UT Southwestern Medical Center, Dallas, TX, USA (e-mail: khu@ustc.edu.cn).

Feng Li, Peng Miao, Zhilei Zhang, Rongqi Sun, and Ge Jin are with State Key Laboratory of Particle Detection and Electronics, University of Science and Technology of China, Hefei, Anhui 230026, P.R. of China (e-mail: phonelee@ustc.edu.cn, mpmp@mail.ustc.edu.cn, zzlei@mail.ustc.edu.cn, srq@mail.ustc.edu.cn, goldjin@ustc.edu.cn).

and state monitoring of FEB can be accomplished with the ADC integrated in GBT-SCA.

### III. DESIGN OF FEB CONFIGURATION TEST BOARD

#### A. Hardware Design

The diagram of the FEB configuration test board is illustrated in Fig.2. The board is divided into the following modules, including FPGA control module, GBT-SCA configuration chip, VMM3 chip, TDS2 chip, network communication module, data interface module, E-link interface and GTX high-speed interface, power supply module, and clock module.

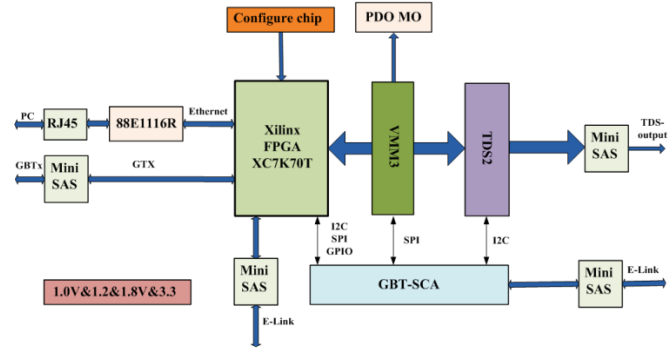

Fig. 2. Diagram of Hardware Design for Configuration Test Board

The signals from sTGC detector are initially received by a VMM3, which is being developed to read out the signals of pad, strip and wire of the sTGC detector. Then, data from VMM3 is processed by the Trigger-Data-Serializer (TDS2) before being transmitted to the Router, and the VMM3's data can be read into the FPGA. The GBT-SCA has the purpose to distribute the control and monitoring signals to VMM3 and TDS2.

The configuration test board uses Xilinx's Kintex 7 (XC7K70T-1FBG484I)[8] high performance and low power FPGA as the main control module of the whole system, the chip provides users with up to 300 IO pins, enough to meet the communication interface needs of our test board, it includes VMM interface, TDS interface, network interface, and E-link interface. FPGA also has high-speed GTX transceivers, two of them are used to connect with the external mini-SAS interface, the high speed output signal of TDS can be connected to FPGA so that the single board can verify the data transmission function of TDS.

VMM has 64 input channels, corresponding to i0-i63, the trigger data directly connected to TDS. The original data of VMM is sent to ROC by DT0 and DT1 after 8b/10b coding (the VMM output of this design is connected to FPGA). In addition, there are also VMM3 configuration SPI bus, as well as some clock and trigger signals. MO(monitor multiplexed analog output), PDO(Peak Detector multiplexed output), TDO(Time detector multiplexed analog output) signals are led to SMA connector for signal monitoring.

TDS has a total of 128 pairs of input interfaces, one TDS can receive two VMMs data output, TDS needs an external 40MHz reference clock, the internal ePLL produces the working clock, the high speed output of TDS is the trigger data of the 4.8Gbps. The configuration of TDS uses I2C bus, which requires only three lines to communicate with external interface.

GBT-SCA has two E-link ports connected to GBTx, actually using only one E-link port, the other is standby. Usually, GBT-SCA is connected to GBTx's special slow control port through E-link. The E-link port runs in 40MHz double data rate(DDR) mode, providing 80Mbps effective data rate. GBT-SCA has 8 SPI communication interfaces, 16 I2C communication interfaces and 32 general IO interfaces as master device to control peripheral device. When GBT-SCA is used as slave device, it has 2 E-link ports to communicate with external master device. Besides, the chip also has JTAG interface, 32 ADCs and 4 DACs.

#### B. Firmware Design

FPGA on the FEB configuration test board is mainly responsible for network communication with the host computer, data communication with GBT-SCA chip, data interface with VMM3, and high-speed data interface with TDS2. Figure 3 is FPGA firmware block diagram of configuration test board.

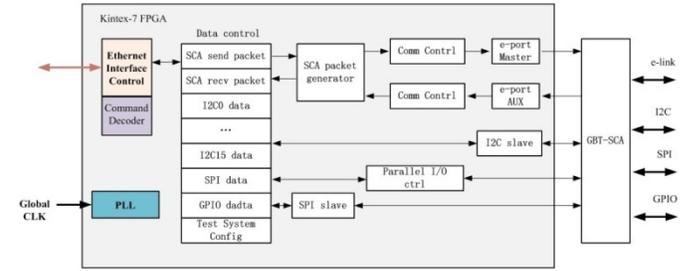

Fig. 3. Block Diagram of FPGA Firmware.

The FPGA firmware includes network transmission module (Ethernet Interface Control/Command Decode), clock module(PLL), data control module (Data control), SCA packet data generation module (SCA packet generator), communication control module (Comm Contrl), E-port module, and GBT-SCA interface.

The network transmission module is responsible for communicating with the host computer. The host computer generates GBT-SCA control instructions, including the generation of GBT-SCA configuration data, the register configuration and control commands of GBT-SCA.

The clock module is used to generate the clock needed for the design logic. The clock of external input 200MHz, through the PLL inside the FPGA generates 125MHz network clock and 40MHz system clock.

The data control module classifies different types of data and protocols, and manages the data in different memory addresses, including GBT-SCA transceiver data, I2C, SPI, GPIO and GBT-SCA configuration data.

SCA packet generator module packages the data in accordance with the HDLC protocol, the communication control module is used to select the GBT-SCA channel (E-port

Master/E-port AUX), GBT-SCA interface includes four types of interface: E-Link, I2C, SPI and GPIO.

The communication between FPGA and GBT-SCA adopts E-link communication protocol, then the configuration data is transferred to VMM3 by GBT-SCA through SPI interface, and to TDS2 by GBT-SCA through I2C interface. Four interfaces of GBT-SCA can be connected to the FPGA, so that the data in FPGA is passed on to GBT-SCA by E-Link, through the GBT-SCA interface output, can be directly returned to FPGA for loopback verification.

## IV. CONFIGURATION TEST

The FEB configuration test board is mainly used to test the GBT-SCA, VMM3, and TDS2 chips on the board, and interactivity with other hardware, including communication interface relationships, such as GBT-SCA communicate with VMM3, GBT-SCA communicate with TDS2, and GBT-SCA communicate with FPGA. The GBT-SCA is the core chip of the configuration test board and researching on controlling of GBT-SCA is the key of FEB configuration test board.

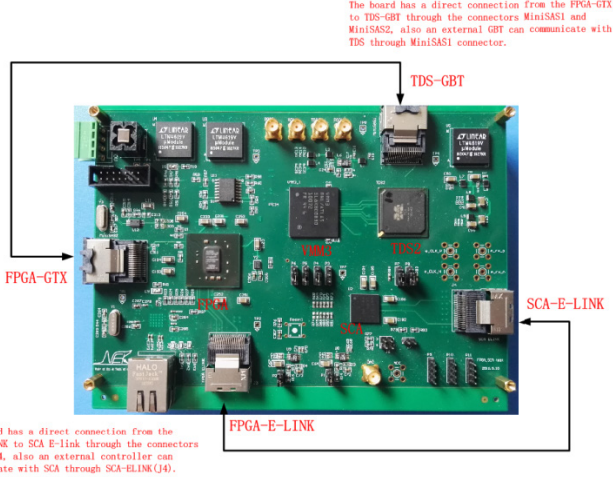

Fig. 4. Test of the Configuration Test Board

The test connection schematic of the FEB configuration test board is shown in Figure 4, the FPGA on the board is used as the main control device. The generation of communication data is controlled by the host computer software, and the communication protocol of the control room is simulated. It is transmitted to FPGA through the network, then, the data of the FPGA is transmitted to GBT-SCA through E-link data transmission protocol. The interface is mini-SAS and connected by mini-SAS cable. GBT-SCA communicate with VMM3 through the SPI interface and communicate with TDS2 through the I2C interface. The output data rate of TDS2 is 4.8Gbps, and the data of TDS2 is transmitted to FPGA for verification, and it is connected with FPGA through the mini-SAS interface. The FEB configuration test board we designed can be independent of the other modules of the system, and the configuration function can be verified by a computer and a FEB configuration board.

### A. VMM3 Configuration Test

VMM3's configuration is completed through the SPI interface, SPI bus consists of 4 signals, respectively ENA, SCK, SDI and CS. When ENA is low and CS is low, VMM3 registers can be accessed through the SCK and SDI, shift with the falling edge of SCK, When CS is high, the configuration data is locked, the register that VMM3 needs to configure is 1728bit, which is written to the internal registers 18 times and 96bits each time, and the configuration timing diagram of 8 VMM3s is shown in Figure 5.

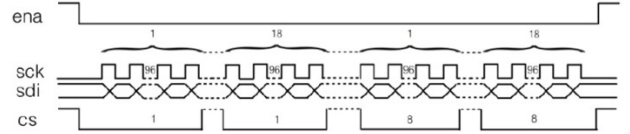

Fig. 5. Configuration Timing Diagram of 8 VMM3s.

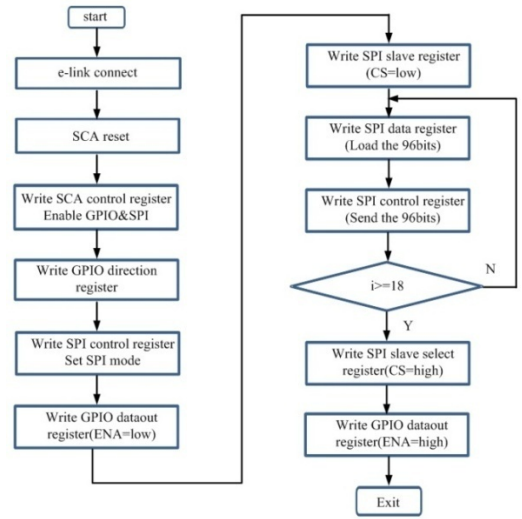

Fig. 6. VMM3 Configuration Flow Chart.

The configuration flow chart is shown in Figure 6. After the software started, the connection instruction is sent to the GBT-SCA, and the E-link connection is established, reset GBT-SCA first, then, write GBT-SCA control registers, and enable GPIO and SPI channels, write GPIO direction register, write SPI control register and set SPI mode, then write GPIO output register(ENA is low), and write SPI slave register(CS is low), so we can load the 96 bits configuration data and write it to SPI data registers, then write SPI control register to send the 96 bits configuration data to VMM3. It need to recycle 18 times to configure a VMM3. When the configuration is completed, the ENA and CS signals will be pulled up, and the configuration will be finished.

### B. TDS2 Configuration Test

The parameters of TDS2 are configured through the I2C interface, and the I2C interface of TDS2 works in the slave mode with 7 bit device addresses, of which 3 bits of the address are used as the address of the multiple TDS2 devices, and the lower 4 bits are used as the address access of the internal registers of the TDS2. There are 16 registers that need to be configured in TDS2, a total of 1296bit, each register varies from 2 bytes to 16 bytes, depending on the configuration task, while GBT-SCA can transmit the most 128 bits data at a time, which is the configuration data of a TDS2 register.

The configuration data of the TDS2 is generated in the host computer, and the host computer sends the configuration data to the FPGA on the configuration test board through the network, then FPGA packages the data into the E-link data protocol in the HDLC format, and sends it to the GBT-SCA. The configuration flow chart is shown in Figure 7. After the software is started, E-link connect instruction is sent to the GBT-SCA. After the connection is set up, the GBT-SCA reset once, the GBT-SCA is initialized, and then the GBT-SCA control register is written to enable the I2C channel, then write the configuration data to the data register. After completing the configuration, we can read out the written data by reading I2C instructions to determine whether it is correct or not.

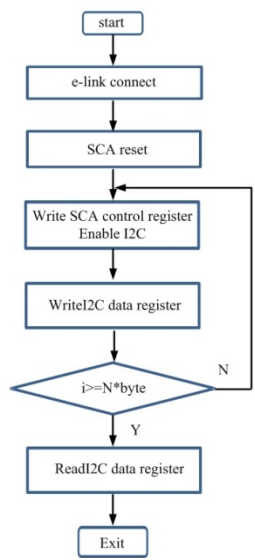

Fig. 7. TDS2 Configuration Flow Chart.

*C. Vertical Slice Test*

Vertical Slice is the integration test of NSW electronics. The main content of the test is the interconnection test of the whole system, determining the functions of interfaces, communications and subsystems. The FEB configuration test board participates in Inter-connection tests at CERN, the configuration of VMM3 and TDS2 can be completed by configuring data link with L1DDC and FELIX, and communication with Pad Trigger Board and Router Board were carried out, as shown in Figure 8 and Figure 9.

V. CONCLUSION

The FEB configuration test board realizes the operation and control of the SCA chip, including E-link, SPI, I2C, GPIO communication, implements the configuration of VMM3 and TDS2, while verifying the TDS2 4.8Gbps high-speed data transfer function. FEB configuration test board also conducted electronic integration test in CERN, it completed communication with Pad trigger and Router board.

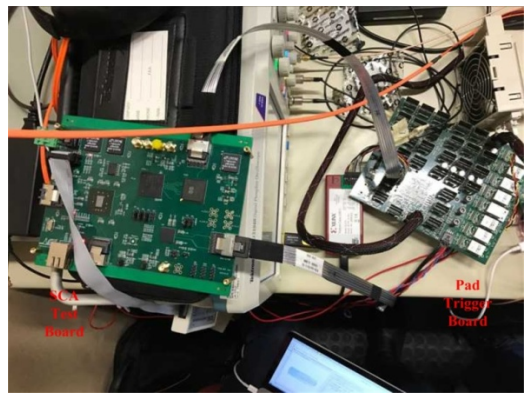

Fig. 8. FEB Configuration Test board and Pad Trigger Board Interconnection Test.

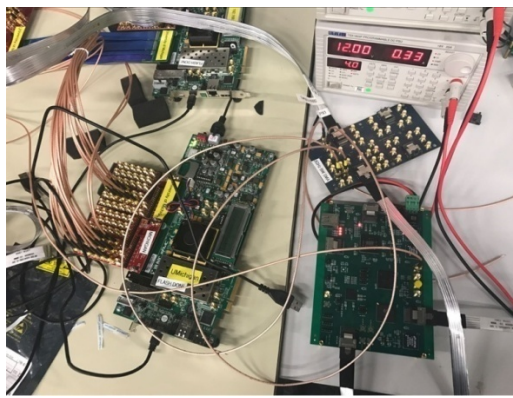

Fig. 9. FEB Configuration Test board and Router Board Interconnection Test.